\documentclass[12pt]{article}
\usepackage{epsfig}
\newcommand{\fnd}[2]{\frac{\textstyle #1}{\textstyle #2}}
\newcommand{\xrm}[1]{{\textstyle \mbox{\rm #1}}}

\begin{document} \baselineskip 6mm
\setcounter{footnote}{2}
\title{Scalar mesons and Adler zeros}
\author{
George Rupp$^{\dag}$, Frieder Kleefeld$^{\dag}$,
and Eef van Beveren$^{\ddag}$ \\[3mm]
$^{\dag\!}${\footnotesize\it Centro de F\'{\i}sica das Interac\c{c}\~{o}es
Fundamentais, Instituto Superior T\'{e}cnico, P-1049-001 Lisboa, Portugal} \\
$^{\ddag\!}${\footnotesize\it Departamento de F\'{\i}sica, Universidade de Coimbra,
P-3004-516 Coimbra, Portugal} \\[4mm]
{\small PACS numbers: 14.40.-n, 11.80.Cr, 12.40.Yx, 13.75.Lb}
}

\date{\today}

\maketitle
\thispagestyle{empty}

\begin{abstract}
A simple unitarized quark-meson model, recently applied with success to light and
charmed scalar mesons, is shown to encompass Adler-type zeros in the amplitude,
due to the use of relativistic kinematics in the scattering sector.
These zeros turn out to be crucial for the description of the $K_0^*$(800)
resonance, as well as the new charmed scalar mesons $D_{s0}^*$(2317)
and $D_0^*$(2300).
\end{abstract}

The light scalar mesons are still causing lots of headaches to many theorists
as well as experimentalists, despite the growing, yet far from general,
consensus that unitarization in some form should represent the necessary setting.
For the purpose of the present short note, we limit ourselves to refer to
a recent brief review \cite{BR04}, which, albeit voicing mainly our personal
view-points, contains many references to work in this field.

An important contribution to unraveling the conundrum of the light scalar
mesons from a data-analysis perspective was recently made by D.~V.~Bugg
\cite{DVB03}, in successfully describing elastic $\pi\pi$ and $K\pi$ scattering
as well as the corresponding $\sigma$ ($f_0$(600)) and $\kappa$ ($K_0^*$(800))
resonances. The clue to this achievement was the inclusion of the respective
Adler zeros for these processes directly into the energy-dependent widths 
contained in the relativistic Breit--Wigner amplitudes used in the phase-shift
fits.

Such $T$-matrix zeros below threshold were determined by S.~Weinberg
\cite{SW66} for elastic $\pi\pi$ scattering, elaborating upon consistency
conditions among strong-interaction amplitudes derived by S.~L.~Adler
\cite{SLA65} for $\pi\pi$, $\pi N$, and $\pi\Lambda$ processes, on the basis
of PCAC. In the case of elastic $\pi\pi$ and $K\pi$ scattering, the Adler
zeros lie at $s^{\pi\pi}_A\approx m^2_\pi/2$ and $s^{K\pi}_A\approx
m^2_K-m^2_\pi/2$, respectively, which were the values used by D.~V.~Bugg.

On the other hand, as early as in 1986 two of us co-authored a paper
\cite{EVB86} in which, for the first time, a complete light nonet of
scalar-meson resonances \cite{MDS82} was predicted with a simultaneous
reasonable description of the corresponding elastic $S$-wave meson-meson phase
shifts. Although the latter coupled-channel model results were obtained without
any fit in the scalar sector, its predictions for the resonance pole positions
lied very close to the present-day world averages. Nevertheless, no attempt was
made in this work to account for dynamical Adler zeros from PCAC or
(approximate)
chiral symmetry, besides the use of a physical input pion mass. Moreover, in a
more recent paper \cite{BR01}, a simplified yet less model-dependent version of
the mentioned coupled-channel approach was employed to fit the $S$-wave $K\pi$
phase shifts up to 1.6 GeV, thereby extracting both the established
$K_0^*$(1430) and the now also listed $K_0^*$(800) resonances \cite{PDG04}.
Finally, we used the very same modified Breit--Wigner formula, with adjusted
quark and meson masses, to successfully describe \cite{BR03} the couple of just
discovered charmed scalar mesons consisting of the very narrow $D^*_{s0}$(2317)
\cite{BABAR03} and the broad $D^*_0$(2300) \cite{BELLE03}.

In view of these surprising results, we analyze here in more detail the
behavior of the latter model amplitude below threshold and in the
complex-momentum plane (see also Ref.~\cite{FK03}). The corresponding
$K^{-1}$ matrix is, for the $1\!\times\!1$ scalar case, simply given by
\begin{equation}
\xrm{cotg}\left(\delta (s)\right)\; =\; \fnd{n_{0}(pa)}{j_{0}(pa)}\; -\;
\left\{ 2\lambda^{2}\mu(s)\,pa j^{2}_{0}(pa)\;\left[\fnd{1.0}{\sqrt{s}-E_1}\;+
\;\fnd{0.2}{\sqrt{s}-E_2}\;-\;1\right]\right\}^{-1} \; ,
\label{bw}
\end{equation}
where $j_0$ and $n_0$ are spherical Bessel and Neumann functions,
respectively, $E_1$ and $E_2$ are the energies of the lowest bare
$J^{PC}\!=\!0^{++}$ $q\bar{q}$ states, $\lambda$ is the coupling for $^3\!P_0$
quark-pair creation, $a$ is the corresponding interaction radius, $p$ is the
relativistic relative momen\-tum of the two-meson system, and $\mu(s)$ is the
associated relativistic reduced mass 
\begin{equation}
\mu(s)\; \equiv\frac{1}{2}\fnd{dp^{2}}{d\sqrt{s}}\; =\;\fnd{\sqrt{s}}{4}\;
\left[ 1\; -\;\left(\fnd{m_{1}^{2}-m_{2}^{2}}{s}\right)^{2}\right]
\; .
\label{mu}
\end{equation}
We immediately see \cite{FK03} from the latter expression that $\mu(s)$
vanishes at $s\!=\!\pm(m^2_1\!-\!m^2_2)$. Therefore, at this point below
threshold, the factor with $\lambda$ in Eq.~(\ref{bw}) is squashed to zero,
so that the amplitude also vanishes. This property was shared by the model of
Ref.~\cite{EVB86}.

Nice illustrations of the effect of these kinematical Adler-type zeros are
the $S$-matrix pole trajectories in the complex-$p$ plane for $S$-wave $DK$
and $D\pi$ scattering (see the Figure). Note that the parametrization here
is slightly different from the one in Ref.~\cite{BR03}, as we now scale
$\lambda$ and $a$ with the reduced mass of the $q\bar{q}$ system, so as to
guarantee rigorous flavor invariance (see Refs.~\cite{BR04}, \cite{FK03} for
details). The result is that, in the $DK$ case, the two pole trajectories
interchange their roles, so that now it is the bare state which gives rise 
to the $D_{s0}^*$(2317) instead of the continuum state. However, despite the
dramatic change in the trajectories themselves,
their end points, corresponding to the value 0.75 (GeV$^{-3/2}$) of the
universal coupling $\lambda_0$, only suffer a modest shift, which even
produces an improved value for the $D_{s0}^*$(2317) mass, viz.\ 2.327 GeV.
In the $D\pi$ case, the trajectories remain qualitatively unaltered, but the
predicted $D_0^*$(2300) pole position also improves, i.e., to
$2.114-i\,0.118$ GeV. However, the crucial message from the Figure appears
to be the large effect of the Adler-type zeros, indicated by the open circles.
Namely, while in the $DK$ case the relatively distant zeros, located at
$p_A\!=\!\pm im_K$, allow the lower pole to travel all the way to the upper
imaginary-$p$ axis, for $D\pi$ scattering the nearby zeros at
$p_A\!=\!\pm i\:\!m_\pi$ slow down this pole so as to settle above threshold.

\noindent\parbox[t]{16.8cm}{
\mbox{ } \\[-2.7cm]
\hspace*{-1.7cm}
\epsfxsize=11.5cm
\epsfbox{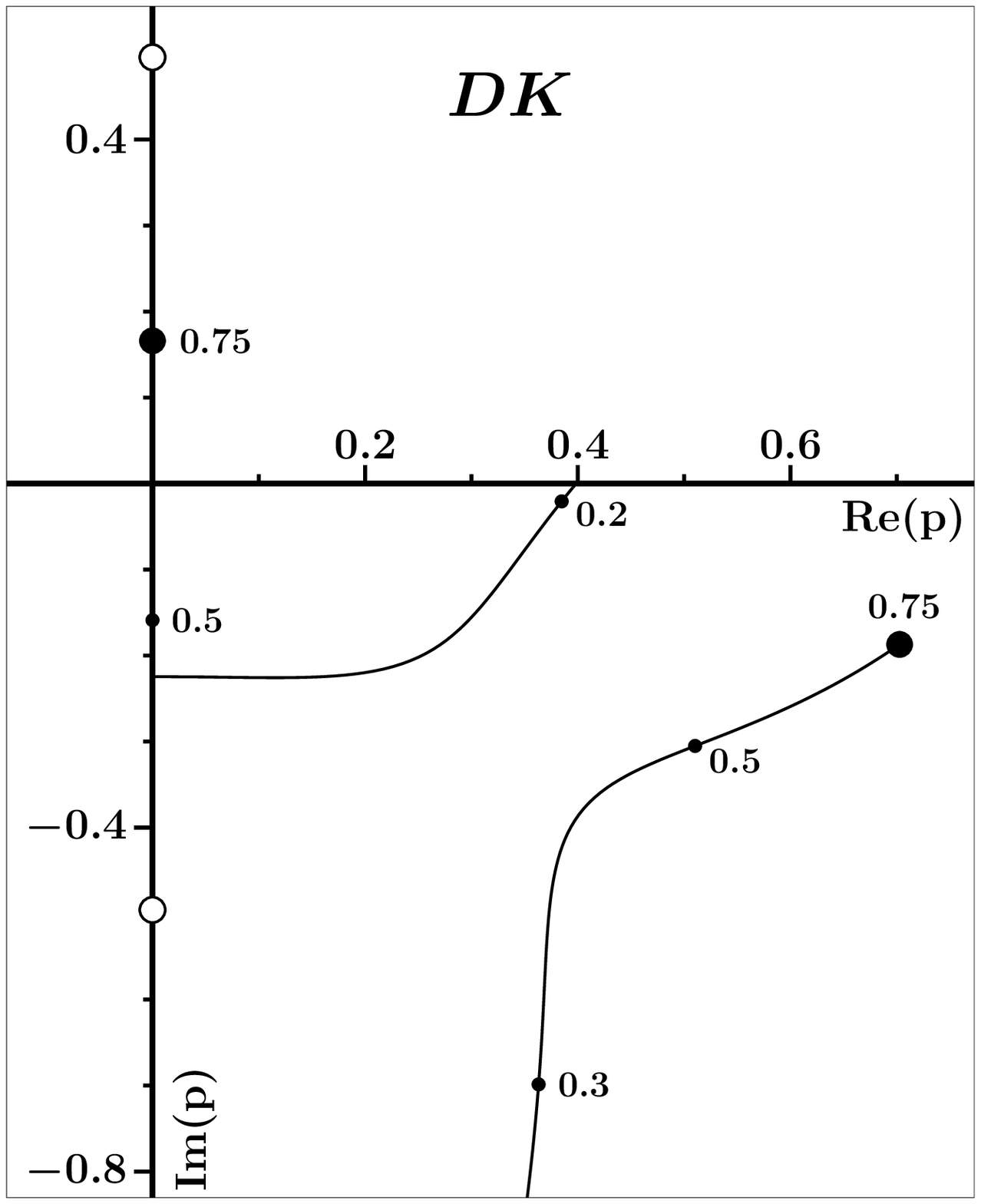}
\hspace{-2.9cm}
\epsfxsize=11.5cm
\epsfbox{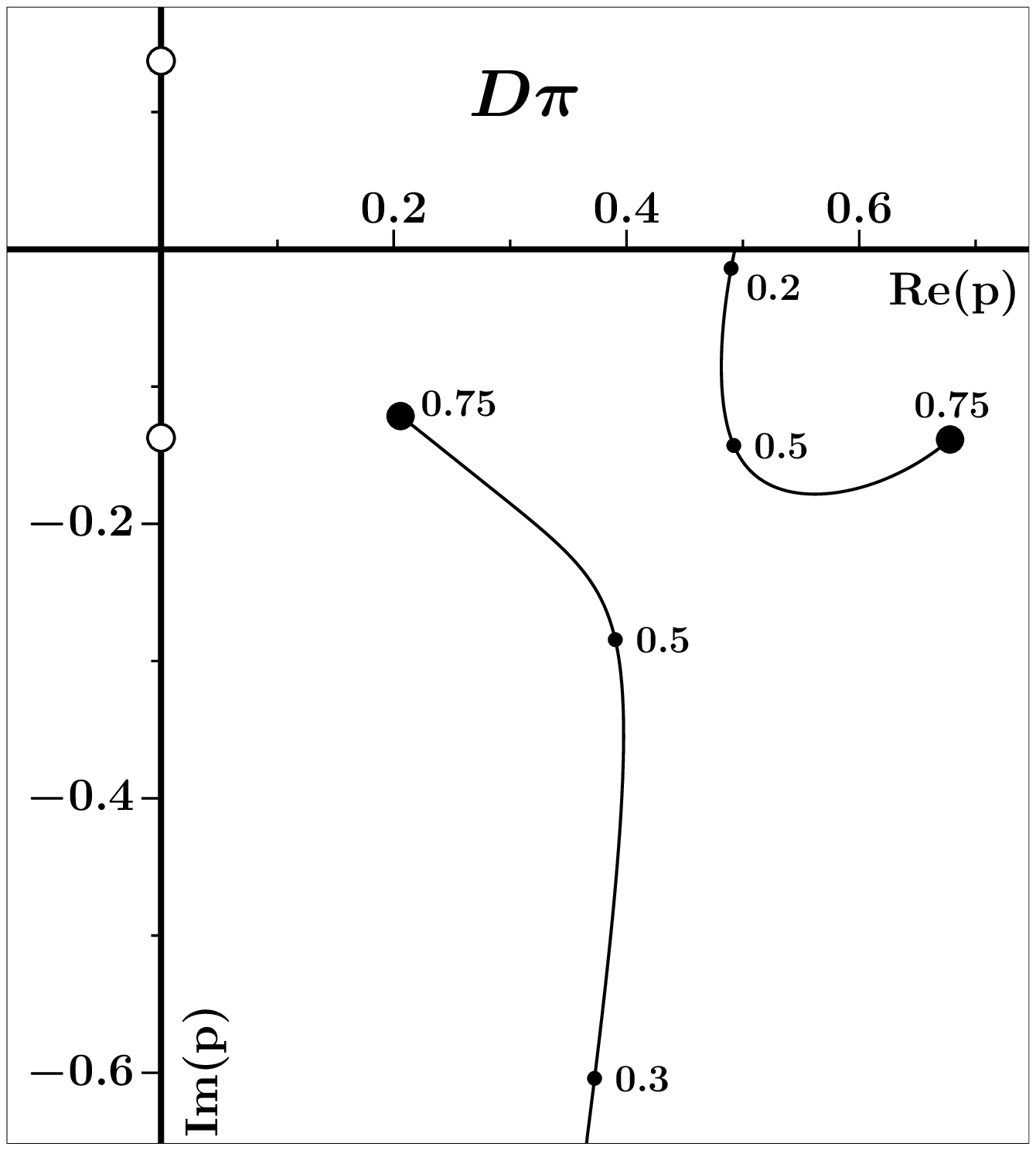}
 \\[-4.0cm]
{\bf FIGURE.} ~Pole trajectories as a function of coupling $\lambda_0$, and
kinematical Adler-type zeros ({$\mathbf \circ$}), in complex-momentum plane
(GeV units), for $S$-wave $DK$ and $D\pi$ scattering. \\[2mm]}

Note that these kinematical zeros deviate less than 1\%  from the theoretical
Adler zeros. Thus, we get a (quasi-)bound state in the former case, and a quite
broad resonance in the latter, similarly to what happens with the $\kappa$ meson
(see Ref.~\cite{FK03} for the $\kappa$-pole trajectory).

As a final test, we check what happens to the $\kappa$ by removing the
Adler-type zero. So we substitute the relativistic reduced mass
(\ref{mu}) by a nonrelativistic one, and then refit $\lambda$, $a$ to the
$K\pi$ phase shifts. The result is a somewhat unphysical \mbox{$\kappa$ pole at}
$E\!=\!483-i\,274$ MeV. Defining also the relative momentum nonrelativistically
then completely kills the $\kappa$, while the $K_0^*$(1430) still survives.
This may provide a clue to the absence of a light $\kappa$ in the analysis of
Ref.~\cite{NAT95}, in which a distant, negative \mbox{Adler zero at}
$s_A\!=\!-0.41$ GeV$^2$ was used, in contrast with e.g.\ the unitarized
chiral approaches of Ref.~\cite{HQZ04}.

\section*{Acknowledgments}
We wish to thank D.~V. Bugg for drawing our attention to the
importance of Adler zeros in scalar-meson data analysis. One of us
(G.R.) also acknowledges useful discussions with N.~A.~T\"{o}rnqvist and
H.~Q.~Zheng. \\
This work was supported by the {\it Funda\c{c}\~{a}o para a
Ci\^{e}ncia e a Tecnologia} \/of the {\it Minist\'{e}rio da
Ci\^{e}ncia, Inova\c{c}\~{a}o e Ensino Superior} \/of Portugal,
under contract no.\ POCTI/\-FNU/\-49555/ 2002 and grant no.\
SFRH/BPD/9480/2002.


\begin{thebibliography}{21}

\bibitem{BR04}
E.~van Beveren and G.~Rupp,
Mod.\ Phys.\ Lett.\ A \textbf{19}, 1949 (2004)
[hep-ph/0406242].

\bibitem{DVB03}
D.~V.~Bugg,
Phys.\ Lett.\ B \textbf{572}, 1 (2003)
[Erratum-ibid. \textbf{595}, 556 (2004)];
Phys.\ Rept.\ \textbf{397}, 257 (2004).

\bibitem{SW66}
S.~Weinberg,
Phys.\ Rev.\ Lett.\ \textbf{17}, 616 (1966).

\bibitem{SLA65}
S.~L.~Adler,
Phys.\ Rev.\ \textbf{137}, B1022 (1965);
\textbf{139}, B1638 (1965).

\bibitem{EVB86}
E.~van Beveren, T.~A.~Rijken, K.~Metzger, C.~Dullemond, G.~Rupp, and
J.~E.~Ribeiro,
Z.\ Phys.\ C \textbf{30}, 615 (1986).

\bibitem{MDS82}
M.~D.~Scadron,
Phys.\ Rev.\ D \textbf{26}, 239 (1982).

\bibitem{BR01}
Eef van Beveren and George Rupp,
Eur.\ Phys.\ J.\ C \textbf{22}, 493 (2001)
[hep-ex/0106077].

\bibitem{PDG04}
S.~Eidelman {\it et al.}  [Particle Data Group Collaboration],
Phys.\ Lett.\ B \textbf{592}, 1 (2004).

\bibitem{BR03}
Eef~van Beveren and George~Rupp,
Phys.\ Rev.\ Lett.\ \textbf{91}, 012003 (2003)
[hep-ph/0305035].

\bibitem{BABAR03}
B.~Aubert {\it et al.} \/[BABAR Collaboration],
Phys.\ Rev.\ Lett.\ \textbf{90}, 242001 (2003)
[hep-ex/0304021].

\bibitem{BELLE03}
K.~Abe {\it et al.} \/[Belle Collaboration],
Phys.\ Rev.\ D \textbf{69}, 112002 (2004)
[hep-ex/0307021].

\bibitem{FK03}
F.~Kleefeld,
AIP Conf. Proc.\ \textbf{717}, 332 (2004)
[hep-ph/0310320].

\bibitem{NAT95}
Nils A. T\"{o}rnqvist,
Z.\ Phys.\ C \textbf{68}, 647 (1995)
[hep-ph/9504372].

\bibitem{HQZ04}
H.~Q.~Zheng, Z.~Y.~Zhou, G.~Y.~Qin, Z.~G.~Xiao, J.~J.~Wang, and N.~Wu,
Nucl.\ Phys.\ A \textbf{733}, 235 (2004)
[hep-ph/0310293];
J.~A.~Oller, E.~Oset, and J.~R.~Pelaez,
Phys. Rev. D \textbf{59}, 074001 (1999)
[Erratum-ibid.\ {\bf 60}, 099906 (1999)]
[hep-ph/9804209].

\end{thebibliography}
\end{document}